\documentclass[9pt,twocolumn,twoside]{optica}
\setboolean{shortarticle}{false}
\setboolean{minireview}{false}
\usepackage{siunitx}
\usepackage{subfig}
\usepackage{comment}
\usepackage{geometry}
\usepackage{xcolor}
\usepackage{graphicx}

\newcommand{\bn}{\textcolor{black}}

\title{Brillouin-based phase shifter in a silicon waveguide}

\author[1]{Luke McKay}
\author[1,*]{Moritz Merklein}
\author[1]{Alvaro Casas Bedoya}
\author[1,+]{Amol Choudhary}
\author[2]{Micah Jenkins}
\author[2]{Charles Middleton}
\author[2]{Alex Cramer}
\author[2]{Joseph Devenport}
\author[2]{Anthony Klee}
\author[2]{Richard DeSalvo}
\author[1]{Benjamin~J.~Eggleton}

\affil[1]{The University of Sydney Nano Institute (Sydney Nano), Institute of Photonics and Optical Science (IPOS), School of Physics, The University of Sydney, 2006, Sydney, NSW Australia}
\affil[2]{Harris Corporation, 1025 W. NASA Boulevard, Melbourne, Florida 32919, United States}
\affil[+]{Current Address: Department of Electrical Engineering, Indian Institute of Technology, Delhi, 110016, India}

\affil[*]{moritz.merklein@sydney.edu.au}

\begin{abstract}
Integrated silicon microwave photonics offers great potential in microwave phase shifter elements, and promises compact and scalable multi-element chips that are free from electromagnetic interference. Stimulated Brillouin scattering, which was recently demonstrated in silicon, is a particularly powerful approach to induce a phase shift due to its inherent flexibility, offering an optically controllable and selective phase shift. However, to date, only moderate amounts of Brillouin gain has been achieved and theoretically this would restrict the phase shift to a few tens of degrees, significantly less than the required \ang{360}. Here, we overcome this limitation with a phase enhancement method using RF interference, showing a \ang{360} broadband phase shifter based on Brillouin scattering in a suspended silicon waveguide. We achieve a full \ang{360} phase-shift over a bandwidth of 15\,GHz using a phase enhancement factor of 25, thereby enabling practical broadband Brillouin phase shifter for beam forming and other applications. 

\end{abstract}

\setboolean{displaycopyright}{true}

\begin{document}

\maketitle

\section{Introduction}
\indent Integrated microwave photonics (IMWP) holds great promise for many applications including radio frequency (RF) phase shifters \bn{and filters}. IMWP benefits from established photonic phase manipulation techniques and has the inherent advantage of broad bandwidths and immunity to RF interference \cite{Marpaung2019,Capmany2006}. RF phase shifting devices are particularly important for phased array antennas (PAAs) which can dynamically change their beam profile by controlling the phase of the signal to each successive antenna element, giving them a high degree of flexibility \cite{doi:10.1036/0071475745,Hansen2009}. PAAs are quickly becoming integral components for many applications such as satellite communication, RADAR, medical imaging, 5G networks and sensing \cite{Naqvi2018,Hansen2009}. \newline
\indent Silicon is the most actively researched material in the field of integrated photonics largely due to its compatibility with complementary metal oxide semiconductor (CMOS) processes, high damage threshold and strong optical nonlinearities \cite{Leuthold2010}. Silicon IMWP offers a massive reduction in size, weight and power consumption (SWaP) allowing a high degree of integration of complex optical structures with low loss on a chip \cite{Jalali:06,Thomson2016,Lin:07,Atabaki2018}.
\begin{figure}[t]
  \centering
   \includegraphics[width=0.5\textwidth]{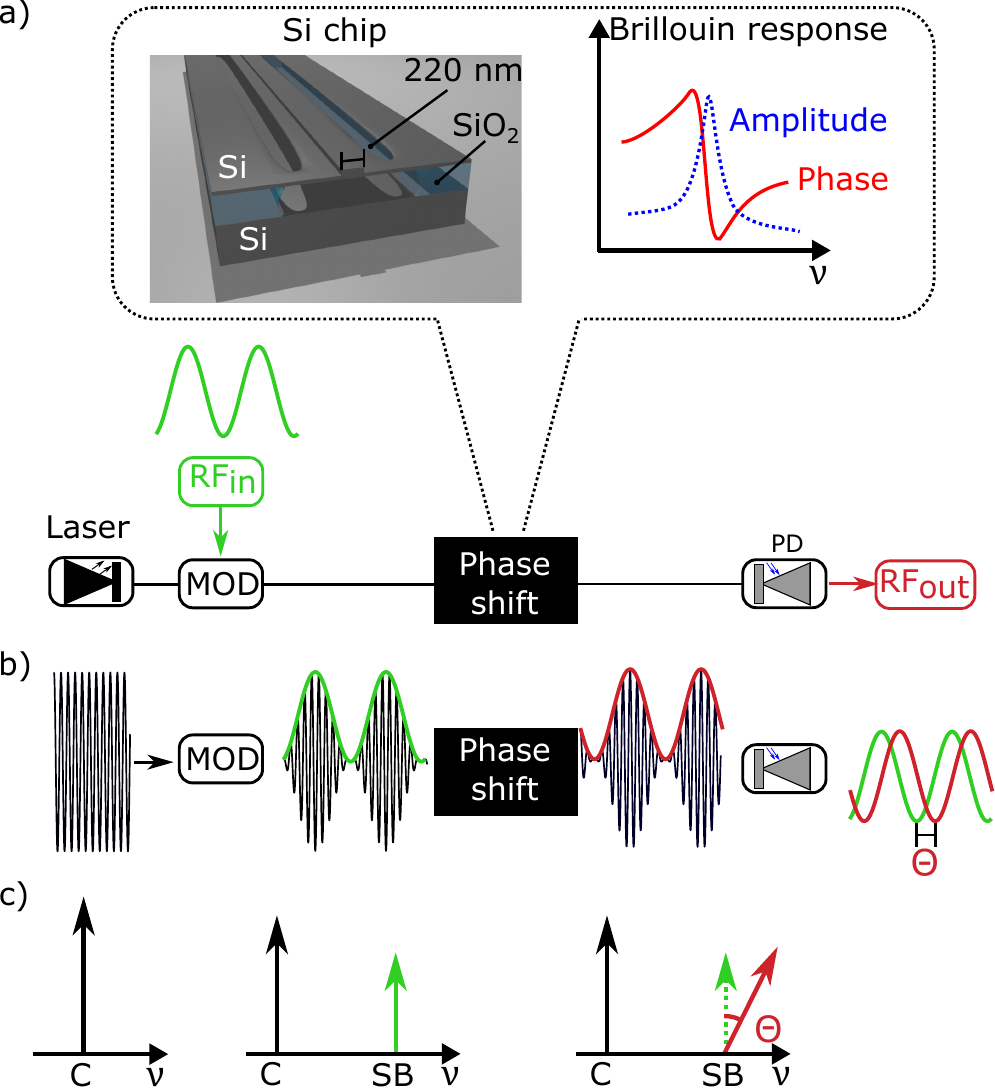}
   \caption{\bn{a) Block diagram of a microwave photonic phase shifter. An input RF signal is modulated onto an optical carrier. The phase of that microwave envelope is shifted in the optical domain using Brillouin scattering in a silicon waveguide (inset) that leads to a phase shifted RF output upon photodetection. b) Time domain representation of an optical carrier modulated by an RF signal and c) the according frequency domain representation. Shifting the phase of the optical sideband results in a phase shifted RF output signal upon beating the optical carrier and sideband at the photodetector. SB, sideband; C, carrier; MOD, modulator; PD, photodetector; \(\nu\), frequency; Si, silicon; SiO\(_2\), silica.}}
  
  \label{overview}
   \hrule
\end{figure}

 The ideal IMWP phase shifter is able to induce a continuously tunable \ang{360} phase shift over a broad bandwidth with minimal insertion loss, small amplitude fluctuations and is integrated on a chip with a small footprint and overall low power consumption. In an IMWP phase shifter, an input RF signal is modulated on an optical carrier creating sidebands. An optical phase shift is then applied to either the optical carrier or sideband, which leads to a phase shift of the RF output signal upon photodetection (Fig. \ref{overview}\,b). There have been many approaches of utilizing an optical phase shift to achieve an RF phase shifter in IMWP systems.
For example, IMWP phase shifters have been realized based on the electro-optic effect utilizing free carrier dispersion and multiplexing/\,de-multiplexing of the optical carrier and sidebands \cite{8379405}. Another way of generating the optical phase shift on a chip is with resonant structures, such as rings and gratings \cite{Burla:14,5200363,5238608}. However, amplitude dependence of the induced phase shift and in the case of rings, limited tunability due to the free-spectral range are challenges for broadband applications. \newline
\indent On the other hand, Brillouin scattering offers a powerful solution as it can induce a tunable, gain-based phase shift over a narrow bandwidth of around 30\,MHz \cite{Merklein2016,Eggleton2013}. More recently, forward Brillouin scattering has been demonstrated in an integrated silicon platform using a suspended waveguide structure, which is capable of guiding both the photons and phonons \cite{Shin2013,Kittlaus2016,SafaviNaeini2019}. 
Microwave photonic phase-shifting schemes based on backward stimulated Brillouin scattering (SBS) have previously been implemented in optical fibre \cite{1561334,Pagani:15}. This concept has been further developed to implement the scheme in an integrated chalcogenide platform, however, due to the limited available on-chip Brillouin gain only a \ang{240} phase shift was achieved \cite{Pagani:14}. To achieve a \ang{360} phase shift, more Brillouin gain would be required, demanding a very large amount of pump power. Even record high Brillouin gain in optimized chalcogenide waveguides of 52\,dB could only achieve a \(\pm\)\ang{160} phase shift \cite{Aryanfar:17}. In silicon, the requirement of ultra-high gain is even more challenging as recent demonstrations of Brillouin scattering in silicon were limited by nonlinear losses \cite{Shin2013,VanLaer2015,Kittlaus2016}. Hence, a technique to enhance the achievable phase response is required. \newline

\indent In this article, we demonstrate a \ang{360} broadband Brillouin-based RF phase shifter in a suspended silicon waveguide with a bandwidth of 15\,GHz. The \ang{360} phase shift was achieved from only 1.6\,dB of Brillouin gain using a phase enhancement factor of 25. Hence, the required pump power is greatly reduced and the RF link gain variation is minimized to only \(\pm\) 3\,dB. The RF bandwidth of this proof-of-principle demonstration was only limited by the available bandwidth of our components and can, in principle, be greatly increased. The broadband phase shift enhancement scheme introduced in this paper is based on RF interference. This can be used for many applications beyond Brillouin-based phase shifters, such as optical signal processing or sensing by reducing power requirements or greatly increasing sensitivity. \newline

\begin{figure}[h]
  \centering
   \includegraphics[width=0.5\textwidth]{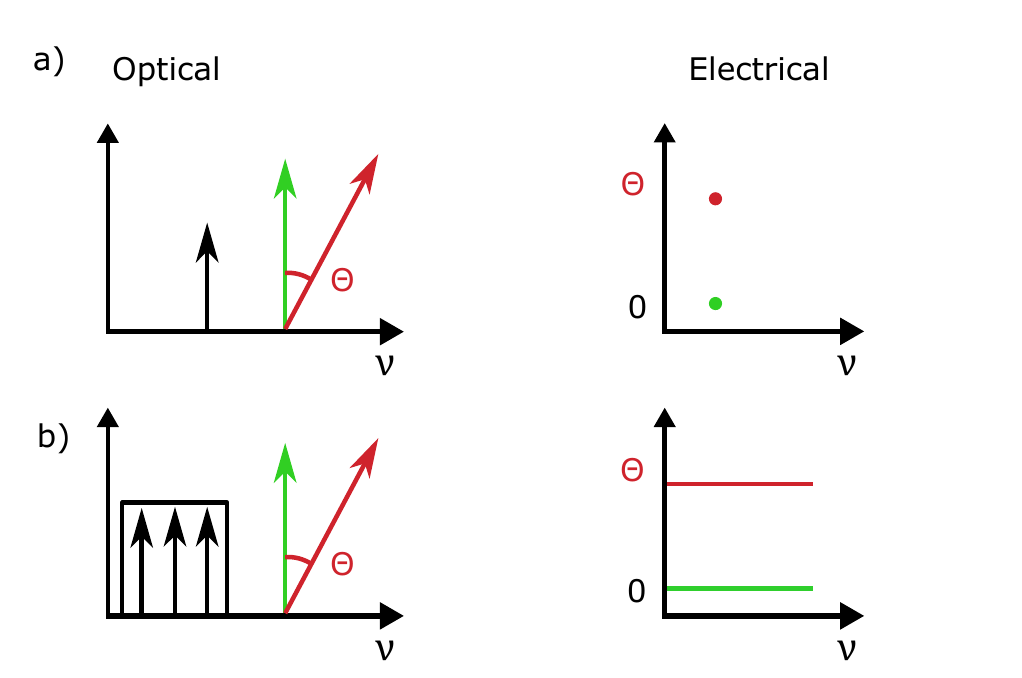}
   \caption{ \bn{a) Narrowband RF tone applied to the optical carrier. The sideband beats with the optical carrier to generate an RF tone in the electrical domain (green). The phase of the RF tone experiences a phase shift of \(\theta\) when a phase shift of \(\theta\) is applied to the carrier. b) A broadband RF signal modulated on the optical carrier. Similarly, the sideband will beat with the carrier to produce a broadband output signal. Similarly, the phase of the broadband signal experiences a phase shift of \(\theta\) when a phase shift of \(\theta\) is applied to the carrier.}}
   \label{broadbandfig}
   \hrule
\end{figure}

\begin{figure*}[ht]
  \centering
   \includegraphics[width=1\textwidth]{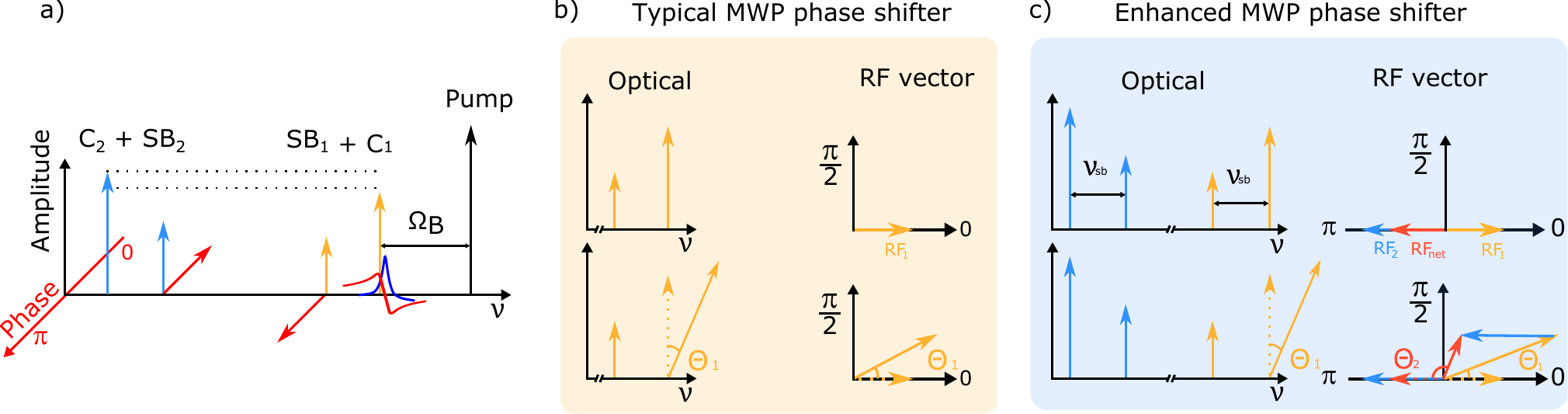}
   \caption{ \bn{a) Basic scheme of the two optical carriers (C\(_1\), C\(_2\)) and sidebands (SB\(_1\), SB\(_2\)) used in the enhanced phase shifter scheme. The lower sideband of C\(_2\) and the upper sideband of C\(_1\) are removed by a filter resulting in out-of-phase RF signals. A Brillouin pump induces an amplification and phase shift of C\(_1\). b) Optical and RF phase space representation of a typical MWP phase shifter. An optical phase shift \(\theta\)\textsubscript{1} applied to one of the sidebands relates directly to a shift \(\theta\)\textsubscript{1} of the RF phase. c) Enhanced MWP phase shifter based on interference and RF vector addition in phase space. As can be seen in the phase space diagram, a small optical phase shift \(\theta\)\textsubscript{1} can lead to a much larger RF phase shift \(\theta\)\textsubscript{2} of the resultant vector RF\(_\mathrm{net}\) (red arrow).}
}
   \label{vectorfig}
   \hrule
\end{figure*}

\section{Principle and Setup}
The core principle of a broadband microwave phase shifter is shown in Fig. \ref{overview}\,a). \bn{An input RF signal (shown in green) is modulated upon an optical carrier. The optical signal propagates through the system where it experiences a phase shift before detection, shown in the time domain in Fig. \ref{overview}\,b). In the frequency domain, the modulation process results in a carrier with optical sidebands and the} phase of the optical sideband is then shifted. This causes a shift in the phase of the RF output signal that is generated at the photodetector by beating the carrier with the sideband (see Fig. \ref{overview}\,c).

In this work, the optical phase shift is induced using Brillouin scattering.  \,\bn{Brillouin scattering involves two co-propagating optical waves interfering to generate an electrostrictive optical force, which in turn generates phonons which are guided in the suspended waveguide structure.  The moving phonons allow one optical signal to coherently amplify and induce a phase shift in the other.} Brillouin scattering is a narrowband \(\chi^{3}\) nonlinear effect that resonantly couples two optical waves - known as Stokes wave \(\omega_{s}\) and a higher frequency pump wave \(\omega_{p}\)-  with an acoustic wave \(\Omega\). This process needs to fulfill energy conservation given by \(\Omega\) =\(\omega_{p}\) - \(\omega_{s}\)  and momentum conservation which can be written as K(\(\Omega\)) = k(\(\omega_{p}\)) - k(\(\omega_{s}\)), where K(\(\Omega\)) is the wavevector of the acoustic wave and k(\(\omega_{p/s}\)) the wavevector of the optical pump and Stokes wave, respectively. While Brillouin scattering can occur in either the forward or backward direction in silicon it is typically studied in forward direction and hence the acoustic wave is mainly a transverse wave and hence the wavevector difference between the pump and Stokes wave is small. \newline
\bn{The optical signal \(\omega_{p}\) coherently amplifies an optical probe that is located at the Stokes frequency \(\omega_{s}\) and in the process is inducing a phase shift as a consequence of the Kramers-Kronig relations \cite{hutchings1992kramers}. This is due to a requirement of causality, which states that a change in amplitude has a corresponding change in refractive index, and hence it experiences a phase shift \cite{agrawal}. The induced optical phase shift due to Brillouin scattering scales linearly with Brillouin gain \cite{Pagani:15}. As we are operating in a low gain regime the Brillouin gain itself is linearly proportional to the effective length and the pump power.} 
\begin{figure*}[h]
  \centering
   \includegraphics[width=1\textwidth]{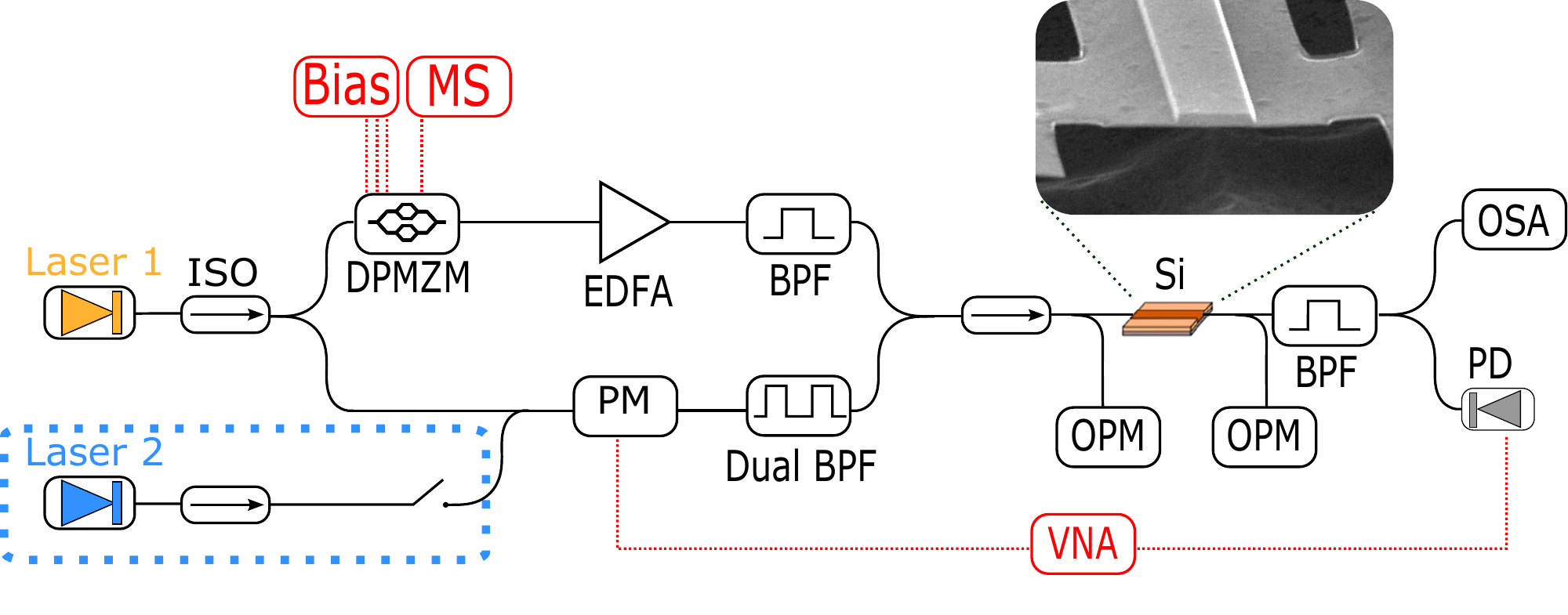}
   \caption{\bn{Schematic of the experimental setup. Laser 1 is split in two arms; one is used as carrier for the input RF signal and the other one acts as the Brillouin pump. Laser 2 is added to achieve the phase enhancement.} Inset: scanning electron microscope (SEM) imagine of a suspended silicon waveguide. ISO: optical isolator; BPF: band-pass filter; DPMZM: dual-parallel Mach-Zehnder modulator; EDFA: Erbium doped fibre amplifier; VNA: vector network analyzer; OPM: optical power meter; OSA: optical spectrum analyzer; PD: photodetector; PM: phase modulator; VCO: voltage controlled oscillator; Si, Silicon chip.}
   \label{simplified}
   \hrule
\end{figure*}
The width of the Brillouin resonance and the according phase shift is given by the inverse of the lifetime \(\tau\) of the acoustic wave and is in our silicon waveguides around 30\,MHz. To achieve a \ang{360} phase shift, however, more than 50\,dB of Brillouin gain is required \cite{Pagani:15}, which would demand an impractical amount of pump power. The maximum achievable gain in silicon is currently far below the threshold required to achieve a \ang{360} phase shift \cite{Kittlaus2016}. \newline
\bn{To achieve a broadband RF phase shift from the narrowband Brillouin phase response, the phase shift is applied to the optical carrier instead of the sideband. A single frequency RF tone is modulated upon the optical carrier and the resulting sideband is separated by the frequency of the RF signal. If no phase shift is applied to the carrier, the resultant RF signal has a phase defined as 0. However, when a phase \(\theta\) is applied to the carrier, the phase of the resultant RF signal will shift to \(\theta\) as shown in Fig. \ref{broadbandfig}\,a). Similarly, this concept expands to the broadband regime where the RF input signal is modulated over a wide frequency range \ref{broadbandfig}\,b). If a phase shift \(\theta\) is then applied to the optical carrier, the entire broadband RF signal will experience a uniform phase shift \(\theta\). So by applying a narrowband phase shift to the carrier, a broadband RF phase shifter can be achieved.} \newline
It has previously been shown that a phase shift can be enhanced using interferometry \cite{Ayun:15,Liu:17,Choudhary2017}. However, so far these schemes have only been applied over a relative narrow bandwidth \cite{Ayun:15,Liu:17,Choudhary2017}. 
Here, on the other hand, we propose and demonstrate a phase enhancement scheme that has a broad bandwidth of several GHz \bn{through acting on the carrier.} \bn{Phase enhancement} is achieved by destructively interfering an out-of-phase RF signal with the original RF signal. \bn{The optical components which generate this interference are shown in Fig. \ref{vectorfig}\,a). The optical carrier from Laser 1 (yellow) and Laser 2 (blue) pass through the same phase modulator, producing sidebands. However, the lower sideband of carrier 2 and the upper sideband of carrier 1 are filtered ensuring the remaining sidebands are out of phase.} This assures that the generated RF signals are out-of-phase when beating on the photodetector.
\bn{In the typical Brillouin phase shifter scheme only laser 1 is connected, the entire phase shift comes from the induced Brillouin phase response as shown in Fig. \ref{vectorfig}\,b). When no gain is present, the resultant signal labeled RF\textsubscript{1} has a phase of 0, but when the pump is on and a phase is applied to the carrier, the resultant phase is \(\theta\).} 
\bn{However, in the phase enhanced scheme,} the RF signal (RF\textsubscript{1}) from carrier 1 destructively interferes with the RF signal from carrier 2 (RF\textsubscript{2}) resulting in a smaller net vector \bn{RF\textsubscript{net}} as shown in Fig. \ref{vectorfig}\,c). The Brillouin phase shift is applied to the carrier of RF\(_{1}\) which was set to be slightly smaller than RF\(_{2}\). Due to the Brillouin amplification and the induced phase shift the ratio between the two RF signals changes and the resultant phase shift can be calculated using vector addition. It is worth noting that the phase shift of the resultant vector \(\theta\)\textsubscript{2} is in the opposite direction to the phase shift experienced by RF\textsubscript{1}, which is given by \(\theta\)\textsubscript{1} in Fig. \ref{vectorfig}\,c). Large phase shifts from only a minimum amount of Brillouin amplification can be achieved when the amplitudes of RF\(_{1}\) and RF\(_{2}\) prior to applying the Brillouin gain are closely matched. \bn{However, due to the destructive interference  inherent in the phase enhancement scheme, there is an unavoidable trade-off between link gain and the phase enhancement factor.} \newline
\begin{figure}[h]
  \centering
   \includegraphics[width=0.5\textwidth]{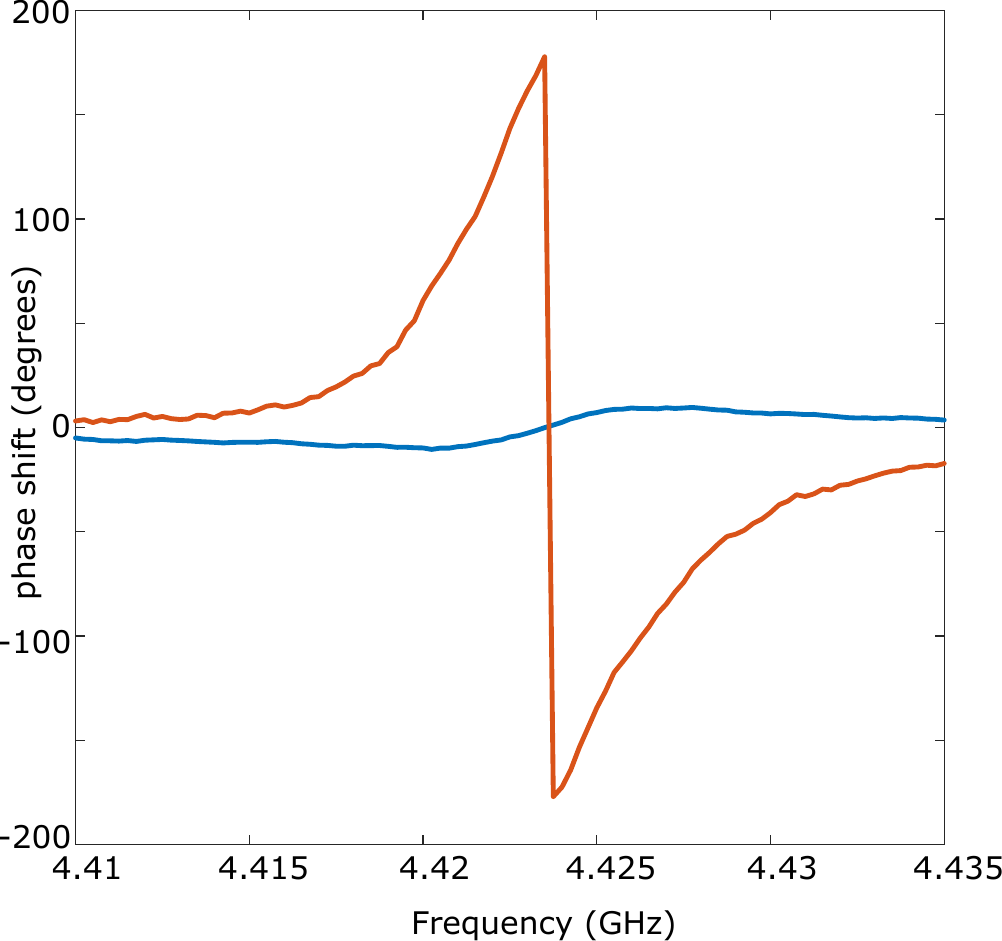}
   \caption{Phase profile without phase enhancement (blue) and with phase enhancement (orange).}
   \label{sideband}
   \hrule
\end{figure}

\begin{figure*}[t]
  \centering
   \includegraphics[width=1\textwidth]{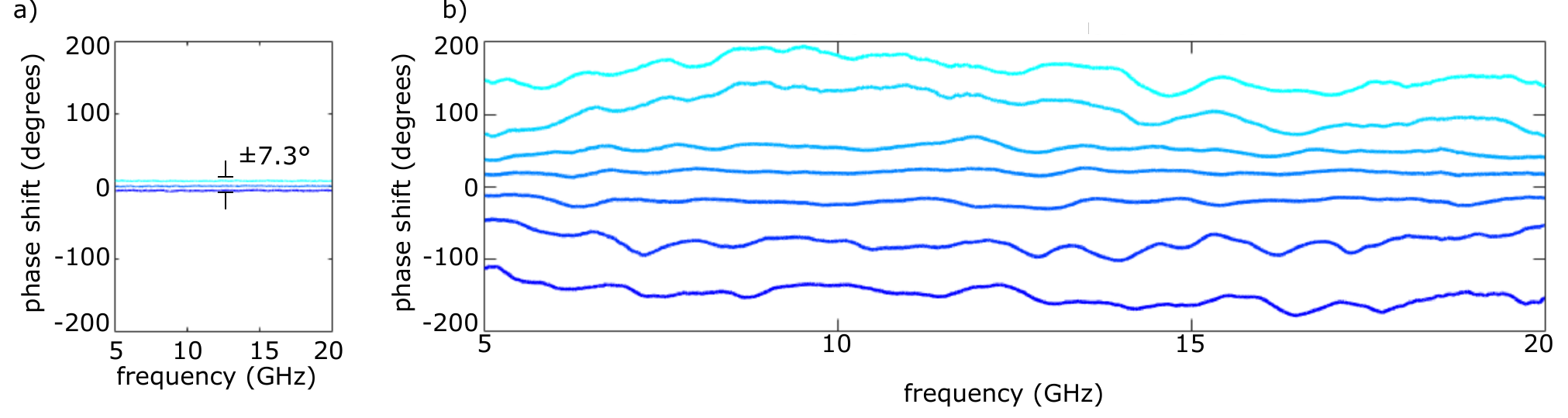}
   \caption{a) Broadband phase shift without phase enhancement. b) Broadband phase shift with an enhancement factor of 25 (a low-pass filter was applied to the data).}
   \label{result}
   \hrule
\end{figure*}
\newpage The experimental setup of our Brillouin RF phase shifter is shown in Fig. \ref{simplified}. A narrow linewidth distributed feedback (DFB) laser (laser 1) is split into two paths using an optical coupler, a pump path and a signal path. The RF input is modulated on the optical carrier in the signal path using a phase modulator, generating two sidebands.  One of these sidebands is then filtered out. The carrier in the pump path is frequency-shifted using a dual-parallel Mach-Zehnder modulator (DPMZM) and is amplified afterwards with an Erbium doped fibre amplifier (EDFA). Both arms are then recombined before they are coupled into the chip using grating couplers. The integrated waveguide in this demonstration is a suspended silicon rib structure with dimensions of  \bn{3050\,nm  wide, 150\,nm thick membrane} with a central ridge protruding \bn{ 70}\,nm with a width of \bn{650}\,nm and a length of 2\,cm. The suspended architecture enables the waveguide to guide photons and phonons without them dissipating into the silica \cite{Kittlaus2016}. \newline 
The waveguide was fabricated at IMEC through ePIXfab. The nanowires were immersed in 10\% diluted hydrofluoric acid with an etching rate of 40\,nm/min for 20 minutes to release them from the silica substrate. The insertion loss of the optical waveguide is 13\,dB, consisting of 3.5\,dB coupling loss per facet using grating couplers and a transmission loss of 3\,dB/cm. \bn{The waveguide used has a slightly lower gain coefficient than what was achieved in previous demonstrations \cite{Kittlaus2016}. Our waveguide structure has a reduced footprint as it is arranged in a spiral but also shows larger losses which limited the achievable net gain. Furthermore, inhomogeneous broadening might limit the gain \cite{Wolff2016}, even though it has been shown to be less severe in suspended ridge waveguides compared to nanowires \cite{Kittlaus2016,VanLaer2015}.} \newline
\indent After the waveguide a highly selective band-pass filter is used to remove the optical pump. The optical carrier and sideband are detected at the photodetector. To achieve the phase enhancement a second laser is coupled into the same phase modulator as laser 1 (see dotted green box in Fig.\ref{simplified}). After the phase modulator, a dual band-pass filter is used to selectively remove the opposite sideband from each optical carrier, so the resulting RF tones are \ang{180} out of phase.
\section{Experimental results}
We first measure the phase response of the forward Brillouin resonance in the silicon waveguide (see Fig. \ref{sideband} shown in blue). We use a vector network analyser (VNA) to sweep an optical probe signal through the Brillouin resonance that is measured to be 4.42\,GHz. The linewidth of the resonance is around 30\,MHz and we can see a \(\pm\)\ang{10} of phase shift for around 2\,dB of Brillouin gain. This moderate amount of phase shift can be enhanced by adding the second, out-of-phase RF signal via the second laser (Fig. \ref{sideband} shown in orange). With the second laser and the vectorial addition of the two RF components we achieve a full \ang{360} phase response with the same amount of pump power and Brillouin gain. Hence a phase enhancement factor of 18 was achieved in this demonstration. \newline
We now want to implement the Brillouin phase shift in a broadband RF phase shifter configuration. Therefore we apply the Brillouin phase shift on the optical carrier signal. This allows us to sweep the sideband over a wide RF frequency range relative to the carrier while all frequencies experience the same phase shift. Using this technique a flat phase shift, i.e. a minimum dependence of the phase shift on RF frequency, can be achieved over tens of GHz that is only limited by the bandwidth of the electo-optic components such as the modulator and the photodetector. Figure \ref{result}\,a) shows a broadband RF phase shift of \(\pm\) \ang{7.3} over a frequency range from 5\,GHz to 20\,GHz. The phase shift was achieved by applying 1.6\,dB of Brillouin gain to the carrier using 18\,dBm coupled pump power \bn{and an average of -40 dB link gain over a broad bandwidth.} 

\begin{figure}[t!]
  \centering
   \includegraphics[width=0.5\textwidth]{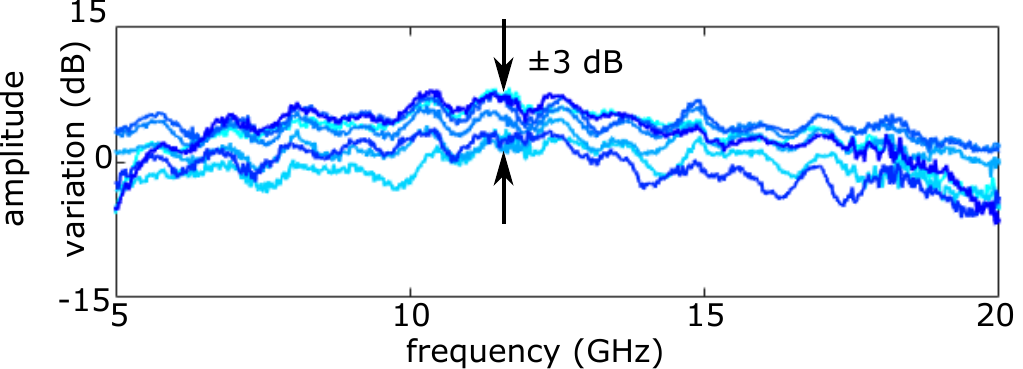}
   \caption{Amplitude variation for a phase shift range of \ang{360} using a factor of 25 phase enhancement. The data is normalized to the underlying RF link. }
   \label{link}
   \hrule
\end{figure}

%\bn{gain was limited due to inhomogeneous broadening, however despite this a \ang{360} phase shift was achieved}. 
The phase shift can be continuously adjusted between \(\pm\) \ang{7.3} by either adjusting the power of the pump to alter the Brillouin gain or by keeping the pump power constant and scanning the pump frequency through the Brillouin resonance. We used the latter approach as scanning the frequency has the advantage of providing direct access to positive and negative phase shift values. In our implementation, the separation of the pump and the carrier is controlled very precisely by the microwave source that is used to shift the pump frequency with the DPMZM. \newline
\indent In the broadband configuration we now add the second laser to show that the phase enhancement scheme also works for broadband RF signals. We therefore keep all the parameters, such as the optical pump power, used in Fig. \ref{result}\,a) but use the phase enhancement to achieve \ang{360} shown in Fig. \ref{result}\,b). Here, both sidebands sweep relative to the carrier from 5\,GHz to 20\,GHz, but in opposite direction, i.e. to lower and higher optical frequencies, respectively. Again, the phase shift is tuned in a continuous way by sweeping through the Brillouin resonance using the microwave signal that drives the DPMZM. The phase enhancement factor in the broadband demonstration was 25 \bn{and the link gain averaged -60 dB over a broad bandwidth. The 20\,dB reduction compared to the non-enhanced case was due to the large phase enhancement factor}. We note that the phase variations with frequency in the phase enhanced case is larger than in the non-enhanced case. One reason is the large phase enhancement factor used in this proof-of-principle demonstration that makes the measurement very sensitive to enviromental fluctuations. A major source for these fluctuations is the strong optical pump wave that co-propagates with the RF signal modulated on the optical wave. This pump wave needs to be filtered out before the photodetector, however, due to the small frequency difference between pump and carrier of less than 5\,GHz, it imposes stringent requirements on the filter. This small frequency spacing means that the filtering allows parts of the pump wave to leak onto the photodetector and also attenuates parts of the carrier and hence, decreases the signal to noise ratio. Also, the roll-off of the filter has a dispersive effect that can cause variations when RF\(_1\) and RF\(_2\) signals interfere. However, these challenges are not fundamental to the scheme and can be overcome by using, for example, a dual waveguide scheme that has the pump in a different waveguide than the carrier signal with an acoustic mode coupling the two modes, as recently demonstrated \cite{Kittlaus2018}. \newline
\indent Another important metric for microwave photonic phase shifter is the amplitude variation with the phase shift. We measured the amplitude variation over the full \ang{360} phase shift in the frequency window from 5\,GHz to 20\,GHz with a phase enhancement factor 25 (see Fig. \ref{link}). The measured variation is \(\pm\)3\,dB. The measured small amplitude variation illustrate another advantage of the phase enhancement scheme as a phase shifter based on Brillouin scattering alone would induce more than 50\,dB (due to the large required gain to achieve a \ang{360} phase shift). Schemes that compensate the link variation by overlying the Brillouin gain with a Brillouin loss response induced by a second laser where put forward in optical fiber to reduce the link variation \cite{1561334,Pagani:14}. While a great reduction of amplitude fluctuations could be achieved the maximum frequency is limited to twice the Brillouin shift of the waveguide. 

\section{Discussion}
It is foreseeable that the system could be implemented entirely in an integrated platform as all the required components have already been demonstrated. Integrating the whole scheme on a chip would not only further reduce size, weight and power (SWaP) consumption but also increase the stability of the interference-based scheme due to smaller fluctuations in relative optical path length. \bn{Furthermore, avoiding coupling to and from the chip would decrease losses, which combined with improved Brillouin gain, would reduce the required enhancement factor and thus increase the overall link gain. In addition, improving the RF amplifier, reducing waveguide losses, increasing filter roll-off to remove the pump before photodetection, or a waveguide with a larger Brillouin frequency shift to ease the filter requirements, would further improve the overall RF link gain.}\newline
\begin{figure}[h]
  \centering
   \includegraphics[width=0.5\textwidth]{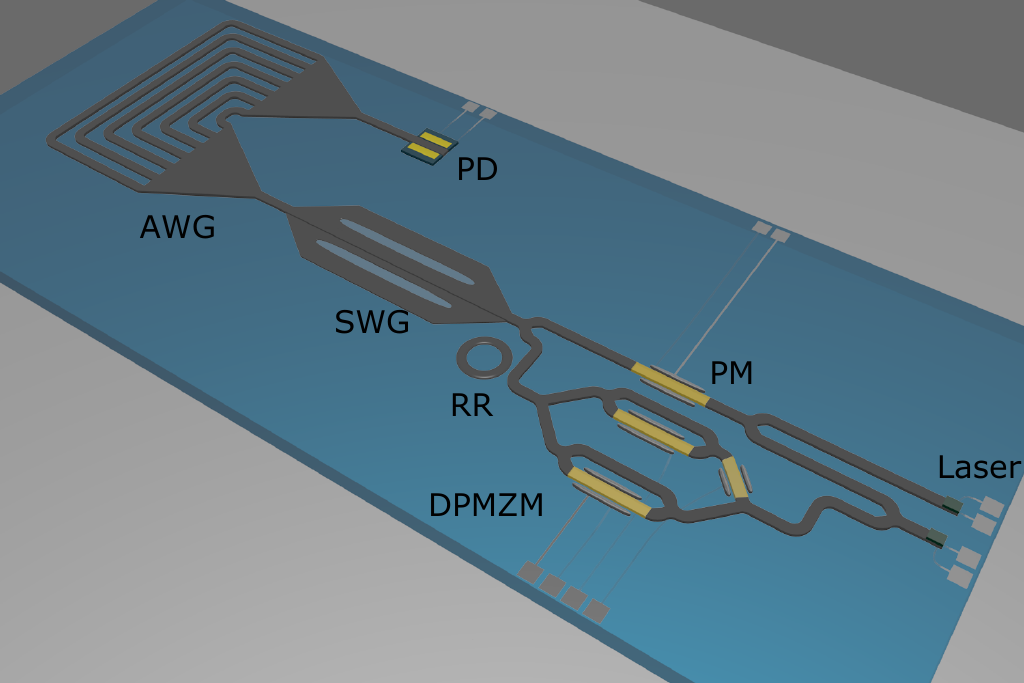}
   \caption{Artists impression of the phase shifting scheme entirely integrated on a chip. AWG, arrayed waveguide grating; SWG, suspended waveguide; PD, photodetector; PM, phase modulator; DPMZM, dual-parallel Mach-Zehnder modulator; RR, ring resonator. }
   \label{concept}
   \hrule
\end{figure}
\indent There are many possible pathways to integrating the whole system on a chip, with the individual devices already showing a high level of development and performance. For example, integrated optical photodetectors have been demonstrated with bandwidths over 100\,GHz and high output power \cite{Beling:16,Li2016}. Also, on-chip modulators have been demonstrated based on different effects, including plasmonic, P-N junctions, and electro-optic materials \cite{Wang2018,Dionne2009,Liu2015}. Laser sources in silicon are still a big challenge due to the indirect bandgap of silicon, however, great progress has been made in III-V based lasers sources \cite{Zhou2015} that can be combined with silicon chips using different bonding techniques, such as flip-chip bonding \cite{Fujioka:10,Tanaka:12}.  Other required components such as integrated optical filters and demultiplexers such as arrayed waveguide gratings (AWGs) are already mature technologies \cite{84502,Kokubun:18}. \newline
\section{conclusion}

Here, we have demonstrated a \ang{360} Brillouin-based RF phase shifter over a bandwidth of 15\,GHz, with only minimal amplitude fluctuations in a CMOS-compatible silicon waveguide. We introduced a broadband RF interference scheme that enhanced the phase shift by a factor of 25, which allowed us to achieve a \ang{360} phase shift from only 1.6\,dB of Brillouin amplification. \bn{The phase enhancement scheme does not only greatly reduce the power requirement, but also enables the first demonstration of an on-chip silicon Brillouin-based phase shifter that achieves a full \ang{360} phase-shift. This is despite silicon waveguides having a significantly reduced absolute Brillouin gain when compared to chalcogenide rib waveguides.} \newline
\bn{A Brillouin-based approach has many advantages over other schemes such as the narrow bandwidth of the phase response, which can be selectively applied to only the carrier. This narrow bandwidth and high selectivity allow the phase shift to be applied at low RF frequencies, where the sidebands are close to the carrier without being effected. Furthermore, one can apply the phase shifter scheme to individual carriers in a wavelength multiplexed scheme. For example, it has been shown that Brillouin scattering can be used to process individual spectral lines of a frequency comb in a single waveguide \cite{Pelusi:17,Choudhary:17}. This would not be possible with inherently broadband phase shifting technologies such as III-V media or thermo-optic phase shifters.} \newline
\indent The broadband phase enhancement concept introduced here is not limited to Brillouin-based phase shifters and can find many other practical applications. It can in general greatly reduce the power requirements of phase shifters, delay lines and true-time delay schemes but can also be utilized in sensing schemes that rely on measuring a phase shift; in the latter, a potentially small phase shift can be enhanced and in the process the sensitivity can be increased.
\section{Acknowledgements}
Australian Research Council (ARC) Linkage grant (LP170100112) with Harris Corporation. U.S. Air Force (USAF) through AFOSR/AOARD (FA2386-16-1-4036); U.S. Office of Naval Research Global (ONRG) (N62909-18-1-2013).
We acknowledge the fabrication of the SOI nanowires, which were done in the framework of the ePIXnet and the University of New SouthWales (UNSW) node of the Australian National Fabrication Facility (ANFF) where the samples were etched.

% Bibliography
\bibliography{sample}

\bibliographyfullrefs{sample}

\end{document}